\documentclass[preprint,tightenlines,showpacs,amsmath,amssymb]{revtex4}

\usepackage{graphicx}
\usepackage{dcolumn}
\usepackage{bm}
\usepackage{rotating}
\usepackage{color}
\usepackage{verbatim} 

\newcommand{\PreserveBackslash}[1]{\let\temp=\\#1\let\\=\temp}
\newcolumntype{C}[1]{>{\PreserveBackslash\centering}p{#1}}
\newcolumntype{R}[1]{>{\PreserveBackslash\raggedleft}p{#1}}
\newcolumntype{L}[1]{>{\PreserveBackslash\raggedright}p{#1}}

\begin{document}
\normalsize
\parskip=5pt plus 1pt minus 1pt

\title{\boldmath Search for the $\eta_c(2S)/h_c\to p\bar{p}$ decays and measurements of the $\chi_{cJ}\to p\bar{p}$ branching fractions}

\author{
M.~Ablikim$^{1}$, M.~N.~Achasov$^{8,a}$, X.~C.~Ai$^{1}$, O.~Albayrak$^{4}$, D.~J.~Ambrose$^{41}$, F.~F.~An$^{1}$, Q.~An$^{42}$, J.~Z.~Bai$^{1}$, R.~Baldini Ferroli$^{19A}$, Y.~Ban$^{28}$
% ...
\\
\vspace{0.2cm}
(BESIII Collaboration)\\
\vspace{0.2cm} {\it
$^{1}$ Institute of High Energy Physics, Beijing 100049, People's Republic of China\\
% ...
$^{4}$ Carnegie Mellon University, Pittsburgh, Pennsylvania 15213, USA\\
% ...
$^{8}$ G.I. Budker Institute of Nuclear Physics SB RAS (BINP), Novosibirsk 630090, Russia\\
% ...
$^{19}$ (A)INFN Laboratori Nazionali di Frascati, I-00044, Frascati, Italy; (B)INFN and University of Perugia, I-06100, Perugia, Italy\\
% ...
$^{28}$ Peking University, Beijing 100871, People's Republic of China\\
% ...
$^{41}$ University of Rochester, Rochester, New York 14627, USA\\
$^{42}$ University of Science and Technology of China, Hefei 230026, People's Republic of China\\
% ...
$^{a}$ Also at the Novosibirsk State University, Novosibirsk, 630090, Russia\\
% ...
}
}
\date{\today}

\begin{abstract}

  Using a sample of $1.06\times10^{8}~\psi(3686)$ events
  collected with the BESIII detector at BEPCII, the decays
  $\eta_{c}(2S)\to p\bar{p}$ and $h_{c}\to p\bar{p}$ are searched for,
  where $\eta_c(2S)$ and $h_c$ are reconstructed in the decay chains
  $\psi(3686)\to\gamma\eta_{c}(2S)$, $\eta_{c}(2S)\to p\bar{p}$ and
  $\psi(3686)\to\pi^{0}h_{c}$, $h_{c}\to p\bar{p}$, respectively. No
  significant signals are observed. The upper limits of the product
  branching fractions are determined to be
  $\mathcal{B}(\psi(3686)\to\gamma\eta_c(2S))\times\mathcal{B}(\eta_{c}(2S)\to
  p\bar{p})<1.4\times10^{-6}$ and
  $\mathcal{B}(\psi(3686)\to\pi^0h_c)\times\mathcal{B}(h_{c}\to
    p\bar{p})<1.3\times10^{-7}$ at the 90\% C.L..  The branching
    fractions for $\chi_{cJ}\to p\bar{p}$ $(J=0,~1,~2)$ are also
    measured to be
    $(24.5\pm0.8\pm1.3,~8.6\pm0.5\pm0.5,~8.4\pm0.5\pm0.5)\times10^{-5}$,
    which are the world's most precise measurements.
\end{abstract}

\pacs{13.25.Gv, 13.40.Hq, 14.40.Pq}

\maketitle

\section{Introduction}
Charmonium has been playing an important role in understanding the
dynamics of QCD. Despite the success of QCD
in many aspects of the strong interaction, the charmonium decay
mechanism remains challenging and presents disagreement between
experimental data and theoretical predictions~\cite{Eich}.

In massless QCD models, the processes
$\eta_c/\chi_{c0}/h_c/\eta_c(2S)\to p\bar{p}$ are forbidden by the
helicity selection rule~\cite{helicity}.
However, the experimental observations of the decays $\eta_c/\chi_{c0}\to
p\bar{p}$~\cite{PDG}, as well as  $h_c$ formed in the $p\bar{p}$
annihilation~\cite{hc_first}, indicate substantial
contributions due to finite masses. These observations have stimulated many theoretical
efforts~\cite{pr2,pr1,pr3}.
In Ref.~\cite{zhaogd}, it is pointed out that
the branching fraction of $\eta_c(2S)\to p\bar{p}$ with respect to
that of $\eta_c\to p\bar{p}$ may serve as a criterion to validate the helicity
conservation theorem, and an anomalous decay in $\eta_c(2S)$ might
imply the existence of a glueball.
For the decay $h_c\to p\bar{p}$, possible large branching fractions are suggested.
Authors of Ref.~\cite{pr2} investigate the long distance contribution via
charmed hadron loops and predict $\mathcal{B}(h_{c}\to p\bar{p})=(1.52-1.93)\times10^{-3}$.
In Ref.~\cite{pr1}, a branching fraction of
${\cal B}(h_c\to p\bar{p}) = (3.2\pm0.5)\times10^{-3}$ is predicted by ``factorizing" the
initial and the final states.

In this paper, we report on a search for $\eta_c(2S)$ and $h_c$ decays
into $p\bar{p}$, where $\eta_c(2S)$ is produced from the $\psi(3686)$
radiative transition, while $h_c$ is produced via the isospin-forbidden process
$\psi(3686)\to\pi^{0}h_{c}$.  In addition, we measure the decays
$\chi_{cJ}\to p\bar{p}$ with J = 0, 1, and 2.  The analysis is based
on an $e^{+}e^{-}$ annihilation sample of $1.06\times10^8$ events
taken at $\sqrt{s}=3.686$~GeV~\cite{psiNum}. A 44 $\textrm{pb}^{-1}$
sample taken at $\sqrt{s}=3.65$ GeV is used to estimate the background
contribution from the continuum processes.

\section{Experiment and data sets}
The BESIII detector, described in detail in Ref.~\cite{BES}, has an
effective geometrical acceptance of 93\% of $4\pi$. A helium-based main
drift chamber (MDC) determines the momentum of charged particles
measured in a 1 T magnetic field with a resolution 0.5\%
at 1 GeV/$c$ (the resolutions mentioned in the paper are rms resolutions).
The energy loss ($d$E$/dx$) is also measured with a resolution better than 6\%.
An electromagnetic calorimeter (EMC) measures energies and positions
of electrons and photons. For 1.0 GeV photons and electrons, the energy
resolution is 2.5\% in the barrel and 5.0\% in the end caps, and the position
resolution is 6 mm in the barrel and 9 mm in the end caps.
A time-of-flight system (TOF) with a time resolution of 80 ps (110 ps)
in the barrel (end cap) is used for particle identification.
A muon chamber based on resistive plate chambers with 2~cm position
resolution provides information for muon identification.

An inclusive Monte Carlo (MC) sample of $1.06\times10^8~\psi(3686)$ events is used for background studies.
The $\psi(3686)$ resonance is produced by the event generator KKMC~\cite{KKMC}, and the decays are generated by EvtGen~\cite{EvtGen} with known branching fractions~\cite{PDG}, while the unmeasured decays are generated according to the Lundcharm model~\cite{lundcharm}. Exclusive signal MC samples are generated to determine the detection efficiency and to optimize selection criteria. The $h_c\to p\bar{p}$ and $\eta_c(2S)\to p\bar{p}$ decays are generated according to phase space distributions, and $\chi_{cJ}\to p\bar{p}$ decays are generated with an angular distribution of protons following the form $1+\alpha\cos^2\theta$ in the $\chi_{cJ}$ helicity frame, where $\alpha$ is taken from measured data. GEANT4 is used to simulate events where the measured detector resolutions are taken into consideration~\cite{geant}.

\section{Event selection and background analysis}
Each charged track is required to have its point of closest approach
to the beam line within 1 cm of the beam line in the radial direction and
within 10 cm from the interaction point along the beam direction and to lie
within the polar angle coverage of the MDC, $|\cos\theta|<0.93$ in
laboratory frame.
 The information from the TOF is used to form a likelihood
$\mathcal{L}_{p}$ ($\mathcal{L}_K/\mathcal{L}_{\pi}$) with a proton
(kaon/pion) hypothesis. To identify a track as a proton, the
likelihood $\mathcal{L}_{p}$ is required to be greater than
$\mathcal{L}_K$ and $\mathcal{L}_{\pi}$.

Photons are reconstructed from isolated showers in the EMC which are
at least 15 (25) degrees away from the proton (antiproton)
candidate. Photon candidates in the barrel ($|\cos\theta|<0.8$) and in
the end cap ($0.86<|\cos\theta|<0.92$) must have an energy of at least
25 MeV. Electromagnetic showers close to the EMC boundaries are poorly
reconstructed and excluded from this analysis. To suppress
electronic noise and energy deposits unrelated to the event, the EMC
timing of the photon candidate must be in coincidence with collision
events $0\leq t\leq14$ (in units of 50 ns).

In the $\psi(3686)\to\gamma\eta_c(2S)/\chi_{cJ}\to \gamma
  p\bar{p}$ and $\psi(3686)\to\pi^0 h_c\to \pi^0
  p\bar{p}\to\gamma\gamma p\bar{p}$ selection, the candidate events
  must have two oppositely charged tracks and at least one or two good
  photons, respectively.  To suppress the nonproton backgrounds in
selecting the $\gamma p\bar{p}$ final states, both tracks are required
to be positively identified as protons, while for the $\gamma\gamma
p\bar{p}$ final states only one track is required to be a proton. A
four-constraint (4C) kinematic fit of $\gamma p\bar{p}$ ($\gamma\gamma
p\bar{p}$) candidates is performed to the total initial four-momentum
of the colliding beams in order to reduce background and to improve
the mass resolution. If more photons than required exist in an event,
the best one(s) is(are) selected by minimizing the $\chi^{2}_{4C}$ of
the 4C kinematic fit. Events with $\chi^2_{4C}<40$ are accepted as
$\gamma p\bar{p}$ ($\gamma\gamma p\bar{p}$) candidates. For
$\gamma\gamma p\bar{p}$ candidates, the invariant mass of the two
selected photons is further required to be in the range 0.11 GeV/$c^2
<M(\gamma\gamma)<$0.15 GeV/$c^2$.

\begin{figure}[htbp]
\begin{center}
\includegraphics[height=2.2in,width = 3.in]{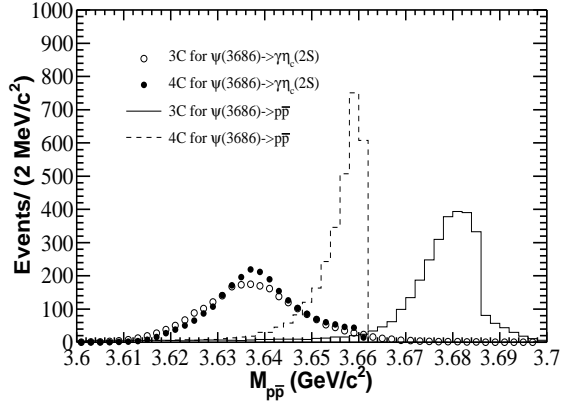}
\caption{Comparison of the invariant mass $M(p\bar{p})$ between 3C and 4C kinematic fits.
For $\psi(3686)\to\gamma\eta_c(2S)$, the open and filled circles are corresponding to 3C and 4C, and for $\psi(3686)\to p\bar{p}$,
the solid and dashed lines are for 3C and 4C, respectively. }
\label{c3_c4_gpp}
\end{center}
\end{figure}

For the $\psi(3686)\to\gamma\eta_c(2S)/\chi_{cJ}\to \gamma p\bar{p}$
channel, the main backgrounds in the $\eta_c(2S)$ signal region
($3.6$ GeV/$c^2$ $\leq M(p\bar{p})\leq3.66$ GeV/$c^2$) are
$\psi(3686)\to p\bar{p}$ decays combined with a fake photon, or with
a photon from initial-state radiation or final-state radiation
(FSR) and the continuum process. In the $\chi_{cJ}$
signal region ($3.3$ GeV/$c^2$ $\leq
M(p\bar{p})\leq3.6$ GeV/$c^2$), the main backgrounds come from the
decays $\psi(3686)\to\pi^0 p\bar{p}$ or the nonresonant process
$\psi(3686)\to \gamma p\bar{p}$.  Since the energy of the transition
photon from $\psi(3686)\to \gamma \eta_{c}(2S)$ is only 50 MeV,
$\psi(3686)\to p\bar{p}$ events can easily fake signal events by combining
with a fake photon. With a 4C kinematic fit, those events will
produce a peak in the $p\bar{p}$ mass spectrum close to the expected
$\eta_c(2S)$ mass.
Therefore, a three-constraint (3C) kinematic fit, where the magnitude of the photon
momentum is allowed to float, is used to determine signal yields. The
3C fit keeps the $\psi(3686)\to p\bar{p}$ peak at the correct position
as the photon momentum tends to zero, and it can separate this
background from the $\eta_c(2S)$ signal efficiently as shown in Fig.~\ref{c3_c4_gpp}~\cite{wangyq}.
The background from the continuum process is studied with the
data taken at $\sqrt{s}=3.65$~GeV. The contribution of the background is
found to be negligible.

\begin{figure}[htbp]
\begin{center}
\includegraphics[height=2.2in,width = 3.in]{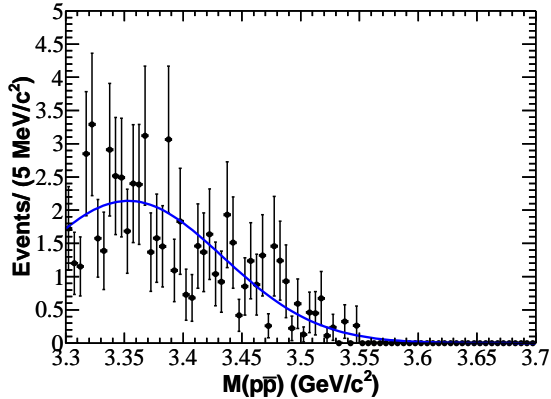}
\caption{The measured background from $\psi(3686)\to\pi^{0}p\bar{p}$ for
the $\psi(3686)\to\gamma p\bar{p}$ mode.
The curve shows the fit with a Novosibirsk function.}
\label{Novo-plot}
\end{center}
\end{figure}

Background from $\psi(3686)\to\pi^0 p\bar{p}$ is measured by selecting
$\pi^0 p\bar{p}$ events from data. The $\pi^0 p\bar{p}$ selection is
the same as that for $\psi(3686)\to\pi^0 h_c$, $h_c\to p\bar{p}$. A MC
sample of $\psi(3686)\to \pi^0 p\bar{p}$ is generated to determine the
efficiencies of the $\gamma p\bar{p}$ selection ($\varepsilon_{\gamma
  p\bar{p}}$) and the $\pi^0 p\bar{p}$ selection ($\varepsilon_{\pi^0
  p\bar{p}}$). The selected $\pi^0 p\bar{p}$ events corrected by the
efficiencies ($\varepsilon_{\gamma p\bar{p}}/\varepsilon_{\pi^0
  p\bar{p}}$) are taken as the $\pi^0 p\bar{p}$ background in
$\psi(3686)\to\gamma p\bar{p}$. The shape of this background can be
described with a Novosibirsk function~\cite{Novo} as shown in
Fig.~\ref{Novo-plot}.

\begin{figure}[htbp]
\begin{center}
\includegraphics[height=2.2in,width = 3.in]{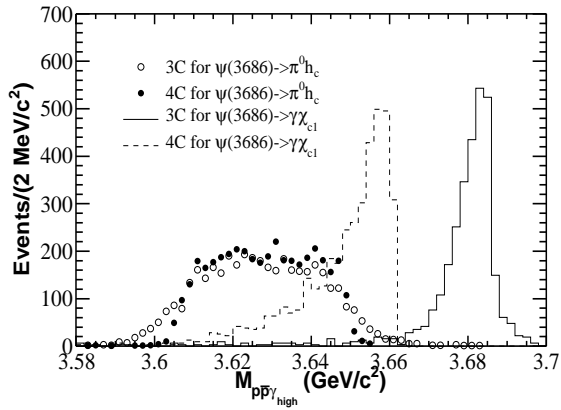}
\caption{Comparison of the invariant mass $M(p\bar{p}\gamma_{\rm high})$ between 3C and 4C kinematic fits.
For $\psi(3686)\to\pi^0 h_c$, the open and filled circles are corresponding to 3C and 4C, and for $\psi(3686)\to\gamma\chi_{c1}$,
the solid and dashed lines are for 3C and 4C, respectively.  }
\label{chi_bkg}
\end{center}
\end{figure}

For $\psi(3686)\to\pi^0 h_c\to\pi^0 p\bar{p}$, the main background
sources are the decays $\psi(3686)\to\gamma\chi_{cJ},\chi_{cJ}\to
p\bar{p}$ (where $J=1,2$) combined with a fake photon
and the $\pi^0 p\bar{p}$ decay from $\psi(3686)$ or continuum process.
The $\chi_{cJ}$ backgrounds are strongly suppressed by using the 3C kinematic fit,
  where the momentum of the photon with lower energy is allowed to float.
  For the $\chi_{cJ}$ backgrounds, the $M(p\bar{p}\gamma_{\rm high})$
  (where $\gamma_{\rm high}$ is the photon with higher energy) with 3C peaks
  at 3.686 GeV/$c^2$, while for the $h_c$ signal, it is below 3.66
  GeV/$c^2$ as shown in Fig.~\ref{chi_bkg}~\cite{wangyq}. A requirement $M(p\bar{p}\gamma_{\rm
    high})<3.66~\rm{GeV}/c^2$ is used to remove this background
  effectively. The $\pi^0 p\bar{p}$ background from the continuum process is studied
  with the data sample taken at $\sqrt{s}=3.65$~GeV and is found not peaking in the signal region.
  The $\pi^0 p\bar{p}$ background
  having the same final state as signal events is irreducible. It is
  included in the fit to the $M(p\bar{p})$ spectrum.

\section{Determination of yields}
\begin{figure}[htbp]
\begin{center}
\includegraphics[height=2.2in,width =3.in]{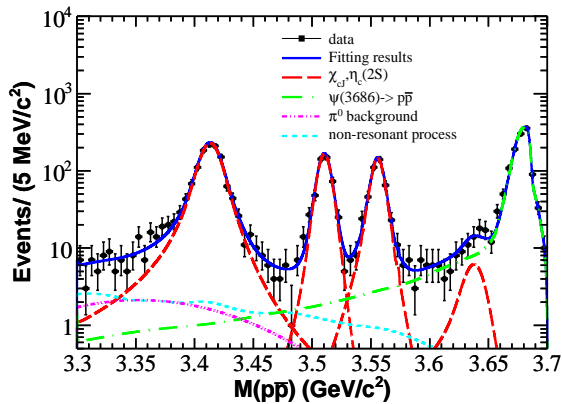}
\caption{The $p\bar{p}$ invariant-mass spectrum after a 3C kinematic
  fit for selected $\psi(3686)\to\gamma p\bar{p}$ candidates from
  data. Dots with error bars are data, the blue solid
    curve is the fitting result, the red long-dashed line is for the
    $\chi_{cJ}$ and $\eta_c(2S)$ signals, the green long-dash-dotted
    line is for $\psi(3686)\to p\bar{p}$, the pink dash-double-dotted
    line is the contribution of $\psi(3686)\to\pi^0 p\bar{p}$ and the
    cyan dashed line is for the non-resonant process.}
\label{fit-etacp}
\end{center}
\end{figure}

Figure~\ref{fit-etacp} shows the $p\bar{p}$ invariant-mass
distribution for the selected $\gamma p\bar{p}$ candidates. There are
clear $\chi_{c0}$, $\chi_{c1}$, $\chi_{c2}$ and $\psi(3686)\to
p\bar{p}$ peaks. The signal for $\eta_c(2S)\to p\bar{p}$ is not
significant. An unbinned maximum likelihood fit to the $M(p\bar{p})$
distribution is used to determine the signal yields of $\eta_c(2S)$ and
$\chi_{cJ}$. The fitting function is composed of signal and background
components, where the signal components include $\eta_c(2S)$ and
$\chi_{cJ}$, and the background components include $\pi^0 p\bar{p}$,
$\psi(3686)\to p\bar{p}$, $\psi(3686)\to \gamma_{\rm FSR}p\bar{p}$ and
nonresonant background. The line shapes for $\eta_c(2S)$ and
$\chi_{cJ}$ are obtained from MC simulation following
$E^{3}_{\gamma}\times BW(m;m_0,\Gamma)\times f_{damp}(E_{\gamma})$,
where $m$ is the invariant mass of $p\bar{p}$, $m_0$ and $\Gamma$ are
the mass and width of the Breit-Wigner line shape for
$\eta_c(2S)$ and $\chi_{cJ}$, and the values are fixed at the nominal
values~\cite{PDG}. $E_{\gamma}$ which equals to
$(m^2_{\psi(3686)}-m^2)/2m_{\psi(3686)}$ is the energy of the
transition photon in the rest frame of $\psi(3686)$, and
$f_{damp}(E_{\gamma})$ is a function that damps the diverging tail
originating from the $E^{3}_{\gamma}$ dependence at the low mass side
(corresponding to high energy of the radiative photon). The form of
the damping factor was introduced by the KEDR collaboration and is
$f_{damp}(E_{\gamma})=\frac{E^{2}_{0}}{E_{\gamma}E_{0}+(E_{\gamma}-E_{0})^{2}}$
~\cite{kedr}, where $E_{0}$ is the peak energy of the transition
photon. The $\pi^0 p\bar{p}$ background is described with a
Novosibirsk function with the fixed shape and amplitude
as described earlier. The backgrounds from $\psi(3686)\to p\bar{p}$
and $\psi(3686)\to p\bar{p}\gamma_{\rm FSR}$ are described with a
shape based on a MC simulation, where the FSR photon is simulated with
PHOTOS~\cite{photo}, and their magnitudes are allowed to float. The
shape of the nonresonant background is determined from a MC
simulation while its magnitude is allowed to float.  To account for a
possible difference in the mass resolution between data and MC
simulation, a smearing Gaussian function $G(\mu,\sigma)$ is convolved
with the line shape of $\chi_{cJ}$, and the parameters of this
function are free in the fit. Since we find that the discrepancy in
the mass resolution decreases with increasing $M(p\bar{p})$ and is
close to zero in the $\eta_c(2S)$ region, a MC-determined line shape
is directly used for the $\eta_c(2S)$ in the fit to data. The fitting results
are shown in Fig.~\ref{fit-etacp}. The signal yields of $\chi_{c0}$,
$\chi_{c1}$, $\chi_{c2}$, and $\eta_c(2S)$ are $1222\pm39$, $453\pm23$,
$405\pm21$ and $34\pm17$, respectively. The statistical significance of the $\eta_c(2S)$
signal is $1.7\sigma$.
The goodness of fit is
$\chi^2/ndf=50.8/65$, which indicates a reasonable fit.

Since $\eta_c(2S)$ signal is not significant, we determine the
upper limit on the number of signal events.
The probability density function (PDF) for the expected
number of signal events is taken to be the likelihood in fitting the
$M(p\bar{p})$ distribution while scanning the number of $\eta_c(2S)$
signal events from zero to a large number, where the signal yields
of the $\chi_{cJ}$ are free.
The 90\% C.L. upper
limit on the number of signal events $N^{\rm up}$, which corresponds to
$\int^{N^{\rm up}}_{0}{\rm PDF}(x)dx/\int^{\infty}_{0}{\rm
  PDF}(x)dx=0.9$, is 54.

\begin{figure}[htbp]
\begin{center}
\includegraphics[height=2.2in,width = 3.in]{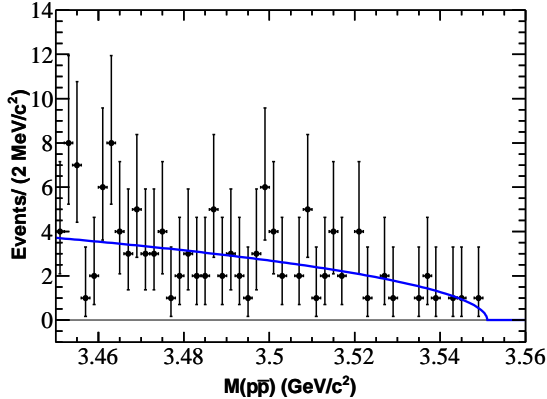}
\caption{The $p\bar{p}$ invariant-mass spectrum for
  $\psi(3686)\to\pi^0 p\bar{p}$. Dots with error bars are data, and
  the blue solid curve is the fitting result with the background
  shape.}
\label{fit-hc}
\end{center}
\end{figure}

Figure~\ref{fit-hc} shows the $p\bar{p}$ invariant-mass distribution
for the selected $\psi(3686)\to\pi^0 p\bar{p}$ candidates. There is no
obvious $h_c\to p\bar{p}$ signal.  The signal yield of $h_c$ is
determined from an unbinned maximum-likelihood fit to the
$M(p\bar{p})$ distribution in $\psi(3686)\to \pi^0 p\bar{p}$ with the
signal and the $\pi^0 p\bar{p}$ background components. The $h_c$
signal is described by the MC determined shape convolved with a
smearing Gaussian. In the MC simulation, the mass and width of $h_{c}$
are set to the measured values~\cite{PDG}. The smearing Gaussian is
used to account for the difference in the mass resolution between data
and MC simulation. The parameters of the Gaussian function are
determined from $\psi(3686)\to\pi^0 J/\psi\to\pi^0 p\bar{p}$. The
$\pi^0 p\bar{p}$ background is described by an ARGUS function~\cite{argus}
with the magnitude and
shape parameters floated. No obvious $h_{c}$ signal event is observed.
The upper limit at the
  90\% C.L. on the $h_{c}\to p\bar{p}$ signal events, calculated with
  the same method as was applied for the $\eta_c(2S)$, is 4.4.
  Figure~\ref{fit-hc} shows the fitting result with the background
  shape, and the goodness of fit is $\chi^2/ndf=18.4/14$.

\section{systematic uncertainties}
In the branching-fraction
measurements, 
there are systematic uncertainties from MDC tracking (1\% per
track)~\cite{eff-track}, particle identification (1\% per
track)~\cite{eff-track}, photon reconstruction (1\% per
photon)~\cite{eff-photon}, the total number of $\psi(3686)$ events
(0.8\%)~\cite{psiNum}, the kinematic fit, and the simulation of
helicity angular distribution of the proton and antiproton. The
uncertainty in the kinematic fit comes from the inconsistency between
the data and MC simulation of the track-helix parameters. We make
corrections to the helix parameters according to the procedure
described in Ref.~\cite{guoyp}, and take the difference between the
efficiencies with and without the correction as the systematic
error. The helicity angular distribution of protons from $\chi_{cJ}$
is taken from measured data and fitted by the formula
$1+\alpha\cos^2\theta$. The $\alpha$ values for
$\chi_{c0},~\chi_{c1}$ and $\chi_{c2}$ are $0.09\pm0.11, 0.12\pm0.20$,
and $-0.26\pm0.17$, respectively. The selection efficiencies are
determined from MC where the $\alpha$ values are set to the mean
values. The change in efficiency by varying the $\alpha$ value by
$\pm1\sigma$ is taken as the uncertainty in the proton angular
distribution. For $\eta_c(2S)/h_c\to p\bar{p}$, the differences in
efficiencies for MC samples simulated with phase space and
$1+\cos^2\theta$, 0.8\% and 0.5\% for $\eta_c(2S)$ and $h_c$,
respectively, are taken as the systematic errors.

For the $\mathcal{B}(\eta_c(2S)/\chi_{cJ}\to p\bar{p})$ measurement,
the uncertainties in the fitting procedure include the damping factor,
fitting range, the description of the $\pi^0$ background,
and the mass resolution of $M(p\bar{p})$.  An alternative
damping function $\exp(-E_{\gamma}^{2}/8\beta^{2})$ was used by CLEO
~\cite{cleo-damp}, where $\beta=65.0\pm2.5$, and $97\pm24$ MeV for
$\eta_c(2S)$ and $\chi_{cJ}$, respectively~\cite{guoyp}.  The
difference in the final results caused by the two damping factors is
taken as the systematic uncertainty. The uncertainty caused by the
fitting range is obtained by varying the limits of the fitting range
by $\pm0.05$~GeV/$c^{2}$.  The uncertainty of the $\pi^0$ background
is estimated by varying the parameters of the shape and magnitude by
$\pm1\sigma$.  The uncertainty from the resolution of
  $M(p\bar{p})$ is found to be negligible.

\begin{table}[htbp]
\begin{center}
\caption{Summary of the relative systematic uncertainties in $\mathcal{B}(\chi_{cJ}\to p\bar{p})$, $\mathcal{B}(\eta_{c}(2S)\to p\bar{p})$ and $\mathcal{B}(h_c\to p\bar{p})$ (in \%).}
\label{sys-etacp}
\begin{tabular}{lC{0.9cm}C{0.9cm}C{0.9cm}C{0.9cm}C{0.9cm}}
  \hline
  Source &  $\chi_{c0}$ & $\chi_{c1}$ & $\chi_{c2}$ & $\eta_{c}(2S)$ & $h_c$ \\
  \hline
  Tracking efficiency        &  2.0   &   2.0   &   2.0   & 2.0  & 2.0\\
  Photon detection           &  1.0   &   1.0   &   1.0   & 1.0  & 2.0\\
  Particle identification    &  2.0   &   2.0   &   2.0   & 2.0  & 1.0\\
  Kinematic fit              &  0.4   &   0.2   &   0.6   & 1.1  & 1.4\\
  Total number of $\psi(3686)$ &  0.8   &   0.8   &   0.8   & 0.8 &0.8 \\
  Damping factor             &  1.1   &   0.1   &   0.2   &  11.8 &-\\
  Fitting range              &  1.4   &   0.4   &   0.2   &  5.9 & 3.4\\
  Background shape        &  0.8   &   0.9   &   0.6   & 8.9  & 12.5\\
  Proton angle distribution  &  0.8   &   0.6   &   0.3   &0.8 & 0.5\\
  Resolution of $M(p\bar{p})$ &- & - & - & - & 5.7 \\
  $\pi^0$ mass region cut & - & - & - & - &  3.0 \\

  Sum & 3.8 & 3.3  & 3.2 & 16.3 & 14.9\\
  \hline
\end{tabular}
\end{center}
\end{table}

For $\mathcal{B}(h_c\to p\bar{p})$, additional uncertainties are caused by the mass resolution
of $M(p\bar{p})$, the fitting range, the $\pi^0$ mass requirement and the background shape. The
uncertainty from the mass resolution of $M(p\bar{p})$ is
estimated by varying the resolution by $\pm1\sigma$. The uncertainty
due to the fitting range is estimated by allowing the fitting range to vary within $0.05$ GeV/$c^2$.
The difference in the number of $h_c$ signal events
is taken as the systematic error. The uncertainty due to the $\pi^0$ mass
requirement is studied using the decay
$\psi(3686)\to\pi^{0}\pi^{0}J/\psi, J/\psi\to l^{+}l^-$
~\cite{bes3-bian}, and 3\% is quoted as the systematic uncertainty. The uncertainty from the background shape (12.5\%)
is estimated by changing the background shape from an ARGUS function to a second-order polynomial.
Table~\ref{sys-etacp} summarizes all the systematic uncertainties. The
overall systematic uncertainties are obtained by summing all the
sources of systematic uncertainties in quadrature, assuming they are
independent.

\section{results and discussion}
We use MC-determined efficiencies to calculate the product branching
fractions
$\mathcal{B}(\psi(3686)\to\gamma\chi_{cJ})\times\mathcal{B}(\chi_{cJ}\to
p\bar{p})$.  By combining the measurements of
$\mathcal{B}(\psi(3686)\to\gamma\chi_{cJ})$~\cite{PDG}, the branching
fractions for $\chi_{cJ}\to p\bar{p}$ are obtained.  The results are
summarized in Table~\ref{branching-chi}.  The upper limits on the
product branching fractions of the $\eta_c(2S)$ and $h_c$ are
calculated with the formula $\frac{N^{\rm up}}{N^{\rm
    tot}\times\varepsilon\times(1-\sigma)}$.  Here $N^{\rm up}$ is the
upper limit of signal events, $N^{\rm tot}$ is the number of
$\psi(3686)$ events, $\varepsilon$ is the MC-determined efficiency 
(45.6\% for $\psi(3686)\to\gamma\eta_c(2S),~\eta_c(2S)\to p\bar{p}$, and
37.7\% for $\psi(3686)\to\pi^0 h_c,~h_c\to p\bar{p}$), and $\sigma$ is
the overall systematic error. We obtain
$\mathcal{B}(\psi(3686)\to\gamma\eta_c(2S))\times\mathcal{B}(\eta_{c}(2S)\to
p\bar{p}) < 1.4\times10^{-6}$ and
$\mathcal{B}(\psi(3686)\to\pi^0
  h_c)\times\mathcal{B}(h_c\to p\bar{p})<1.3\times10^{-7}$ at the 90\%
C.L.. 

\begin{table}[htbp]
\begin{center}
\caption{The selection efficiencies, signal yields extracted from the
  fit, the product branching fractions
  $\mathcal{B}(\psi(3686)\to\gamma\chi_{cJ})\times\mathcal{B}(\chi_{cJ}\to
  p\bar{p})$ and the branching fractions $\mathcal{B}(\chi_{cJ}\to
  p\bar{p})$. Here the first errors are statistical and the second systematic.}
\label{branching-chi}
\begin{tabular}{cC{1cm}C{1.5cm}C{7.0cm}C{4.cm}}
  \hline
  Channels & $\varepsilon(\%)$ & $N_{\rm signal}$ &$\mathcal{B}(\psi(3686)\to\gamma\chi_{cJ})\times\mathcal{B}(\chi_{cJ}\to p\bar{p})(\times10^{-5})$&$\mathcal{B}(\chi_{cJ}\to p\bar{p})(\times10^{-5})$ \\
  \hline
  $\chi_{c0}$ &  48.5 & $1222\pm39$ & $2.37\pm0.08\pm0.09 $ &$24.5\pm0.8\pm1.3$ \\
  $\chi_{c1}$ &  53.8 & $453\pm23$ & $0.79\pm0.04\pm0.03 $ &$8.6\pm0.5\pm0.5$ \\
  $\chi_{c2}$ &  52.0 & $405\pm21$ & $0.73\pm0.04\pm0.03 $ &$8.4\pm0.5\pm0.5$ \\
  \hline
\end{tabular}
\end{center}
\end{table}

The branching fraction for $\eta_c(2S) \to p\bar{p}$ is determined by
multiplying the ratio of the product branching fractions
$\frac{\mathcal{B}(\psi(3686)\to \gamma\eta_c(2S))\times
  \mathcal{B}(\eta_c(2S)\to p\bar{p})}{ \mathcal{B}(\psi(3686)\to
  \gamma\eta_c(2S))\times \mathcal{B}(\eta_c(2S)\to K \bar{K} \pi)}$
and $\mathcal{B}(\eta_c(2S) \to {K\bar{K}\pi})$.  Here the product
branching fraction $\mathcal{B}(\psi(3686)\to \gamma\eta_c(2S))\times
\mathcal{B}(\eta_c(2S)\to K \bar{K} \pi)$ is taken from the recent
BESIII measurement~\cite{bes3-wangll}, and $\mathcal{B}(\eta_c(2S) \to
{K\bar{K}\pi})$ was measured by BABAR~\cite{BaBar}. This allows some
systematic errors, such as errors in the tracking efficiency and the
damping factor, to cancel out.  The result is inflated by a factor
$1/(1-\sigma)$, where the fractional systematic error $\sigma$ is dominated by
the $\mathcal{B}(\eta_c(2S) \to {K\bar{K}\pi})$ measurement.  The 90\%
C.L. upper limit is determined to be $\mathcal{B}(\eta_c(2S)\to
p\bar{p})<4.8\times 10^{-3}$.
By combining the BESIII measurement of
$\mathcal{B}(\psi(3686)\to\pi^{0}h_{c})$~\cite{bes3-bian}, the upper limit of the branching fraction is obtained to be $\mathcal{B}(h_c\to
p\bar{p})<1.7\times10^{-4}$ at the 90\% C.L., where the
errors are treated with the same method as in $\mathcal{B}(\eta_c(2S)\to p\bar{p})$.

In summary, with a sample of $1.06\times10^8~\psi(3686)$ events, we
search for the decays $\eta_c(2S)\to p\bar{p}$ and $h_c\to p\bar{p}$,
but no significant signals are observed. The 90\% C.L. upper limits of
the branching fractions for $\eta_c(2S)\to p\bar{p}$ and $h_c\to
p\bar{p}$ are determined.  The current upper limit of
$\mathcal{B}(\eta_c(2S)\to p\bar{p})$, which is larger than the
measurement of $\mathcal{B}(\eta_c\to p\bar{p})$~\cite{bes3-guoaq},
cannot directly test the conjecture of Ref.~\cite{zhaogd} to
validate the helicity theorem.
The upper limit on $\mathcal{B}(h_c\to p\bar{p})$ obtained from this
work is consistent with the earlier experimental
results~\cite{hc_first} and is lower than the
predictions~\cite{pr1,pr2}, where model parameters may
need to be tuned.  The branching fractions of $\chi_{cJ}\to p\bar{p}$
are measured with improved precision, consistent with the most recent
measurement by CLEO-c~\cite{ppbar-cleo}, and the results are also
compatible with theoretical calculation of $\mathcal{B}(\chi_{cJ}\to
p\bar{p})~(J=0,1,2)$ by including the color octet
contribution~\cite{com}. The results presented in this paper will be
of interest for future experiments like PANDA in their search for
hadronic resonances~\cite{panda}.

\section{ACKNOWLEDGMENTS}
The BESIII collaboration thanks the staff of BEPCII and the computing center for their strong support. This work is supported in part by the Ministry of Science and Technology of China under Contract No. 2009CB825200; National Natural Science Foundation of China (NSFC) under Contracts No. 10625524, No. 10821063, No. 10825524, No. 10835001, No. 10935007, No. 11125525, and No. 11235011; Joint Funds of the National Natural Science Foundation of China under Contracts No. 11079008 and No. 11179007; the Chinese Academy of Sciences (CAS) Large-Scale Scientific Facility Program; CAS under Contracts No. KJCX2-YW-N29 and No. KJCX2-YW-N45; 100 Talents Program of CAS; German Research Foundation DFG under Collaborative Research Center Contract No. CRC-1044; Istituto Nazionale di Fisica Nucleare, Italy; Ministry of Development of Turkey under Contract No. DPT2006K-120470; U. S. Department of Energy under Contracts No. DE-FG02-04ER41291, No. DE-FG02-05ER41374, No. DE-FG02-94ER40823, and No. DESC0010118; U.S. National Science Foundation; University of Groningen (RuG) and the Helmholtzzentrum fuer Schwerionenforschung GmbH (GSI), Darmstadt; and WCU Program of National Research Foundation of Korea under Contract No. R32-2008-000-10155-0.



\begin{thebibliography}{99}
\bibitem{Novo}
The Novosibirsk function is defined as $f(m_{ES})=A_{S}{\rm exp}(-0.5\ln^{2}[1+\Lambda\tau\cdot(m_{ES}-m_0)]/\tau^2+\tau^2)$, where $\Lambda=\sinh(\tau\sqrt{\ln4})/(\sigma\tau\sqrt{\ln4})$, the peak position is $m_0$, the width is $\sigma$, and $\tau$ is the tail parameter.
\end{thebibliography}
\end{document}